\documentclass[pra,showpacs,twocolumn]{revtex4}
\usepackage{graphicx}

\begin{document}

\title{Radiation-pressure cooling and optomechanical instability of a micro-mirror}

\author{O. Arcizet}
\author{P.-F. Cohadon}
\author{T. Briant}
\author{M. Pinard}
\author{A. Heidmann}
\affiliation{Laboratoire Kastler Brossel, Case 74, 4 place
Jussieu, 75252 Paris Cedex 05, France }

\pagestyle{plain}
\maketitle

Recent experimental progress in table-top experiments
\cite{Rugar,EurophysLetters} or gravitational-wave interferometers
\cite{LIGO} has enlightened the unique displacement sensitivity
offered by optical interferometry. As the mirrors move in response
to radiation pressure, higher power operation, though crucial for
further sensitivity enhancement, will however increase quantum
effects of radiation pressure, or even jeopardize the stable
operation of the detuned cavities proposed for next-generation
interferometers \cite{Meers,Heinzel,CavityDetuning}. The
appearance of such optomechanical instabilities
\cite{Vinet,Instability} is the result of the nonlinear interplay
between the motion of the mirrors and the optical field dynamics.
In a detuned cavity indeed, the displacements of the mirror are
coupled to intensity fluctuations, which modifies the effective
dynamics of the mirror. Such "optical spring" effects have already
been demonstrated on the mechanical damping of an electromagnetic
waveguide with a moving wall \cite{Braginsky}, on the resonance
frequency of a specially designed flexure oscillator
\cite{Australiens}, and through the optomechanical
 instability of a silica micro-toroidal resonator \cite{Vahala}.
  We present here an experiment
where a micro-mechanical resonator is used as a mirror in a very
high-finesse optical cavity and its displacements monitored with
an unprecedented sensitivity. By detuning the cavity, we have
observed a drastic cooling of the micro-resonator by intracavity
radiation pressure, down to an effective temperature of
$10\,\mathrm{K}$. We have also obtained an efficient heating for
an opposite detuning, up to the observation of a
radiation-pressure induced instability of the resonator. Further
experimental progress and cryogenic operation may lead to the
experimental observation of the quantum ground state of a
mechanical resonator \cite{Roukes,Cleland,Schwab}, either by
passive \cite{Karrai} or active cooling techniques
\cite{LaserCool,Article1,SpieAustin}. \bigskip

\begin{figure}[b]
\includegraphics[width=6cm]{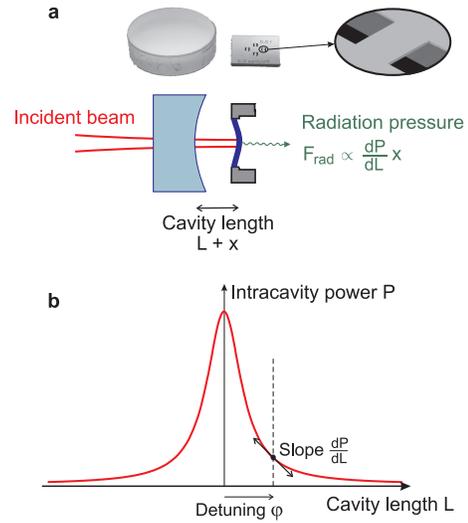}
\caption{Principle of radiation-pressure cooling. {\bf a}, Layout
 of the optical cavity. The micro-resonator mirror is etched
upon a $1\,\mathrm{cm}^2$ silicon chip. The coupling mirror of the
cavity is a standard low-loss silica mirror. {\bf b}, Intracavity
power $P$ as a function of the cavity length $L$. Around a working
point defined by the normalized cavity detuning $\varphi$, the
force $F_{\mathrm{rad}}$ exerted by the intracavity field upon the
micro-resonator is proportional to the slope of the Airy peak and
to the displacement $x$ of the resonator.} \label{fig:Phase}
\end{figure}

The resonator is placed at the end of a linear cavity, along with
a conventional coupling mirror (Fig. 1a). As we are only
interested in the motion at frequencies $\Omega$ close to a
resonance frequency $\Omega_\mathrm{m}$ of the resonator, the
mirror dynamics can be approximated as the one of a single
harmonic oscillator, with resonance frequency
$\Omega_{\mathrm{m}}$, mass $M$, damping $\Gamma_{\mathrm{m}}$ and
mechanical susceptibility:
 \begin{equation}
\chi_{\mathrm{m}}\left[\Omega\right]=\frac{1}{M\left(\Omega_{\mathrm{m}}^2-\Omega^2-i\Gamma_{\mathrm{m}}\Omega\right)}.
\end{equation}
The resonator is submitted to a radiation pressure force
$F_{\mathrm{rad}}$ induced by the intracavity field. Depending on
the detuning $\Psi\equiv 4\pi L/\lambda\,[2\pi]$, where $L$ is the
cavity length and $\lambda$ the laser wavelength, any small
displacement $x$ of the resonator induces a variation of the
intracavity power $P$ and of the radiation pressure (see Fig. 1b).
As a consequence, the spring constant $k=M\Omega_{\mathrm{m}}^2$
of the resonator is balanced by the radiation pressure force: for
a positive detuning, the displacement creates a negative linear
force and thereby an additional binding force, increasing the
effective spring constant, whereas for a negative detuning,
 the force corresponds to a softening of the oscillator.
  Effects are null at resonance,
maximum at half-width of the optical resonance, and proportional
to the incident power. These effects have already been directly
observed \cite{Australiens}, as well as the bistable behaviour of
the cavity for a negative detuning \cite{Dorsel}. In our
experiment, however, due to the high-finesse cavity and the high
resonance frequency, such a static approach is no longer valid and
one has to take into account the cavity storage time to evaluate
the resonator dynamics. Additional dephasings appear and the
radiation pressure force $F_{\mathrm{rad}}$ can be written in
Fourier space \cite{CavityDetuning}:
\begin{equation}
F_\mathrm{rad}\left[\Omega\right]=-2\frac{\varphi\varphi_{\mathrm{NL}}}{\Delta}M\Omega_{\mathrm{m}}^2
\,x\left[\Omega\right],\label{eq:Frad}
\end{equation}
with
\begin{equation}
\Delta=\left(1-i\Omega/\Omega_{\mathrm{c}}\right)^2+\varphi^2,
\label{eq:Delta}
\end{equation}
 where $x\left[\Omega\right]$ is the Fourier component of the
resonator displacement, $\varphi=\Psi/\gamma$ the detuning
normalized to the cavity damping rate $\gamma$, and
$\Omega_{\mathrm{c}}$ the cavity bandwidth.
$\varphi_{\mathrm{NL}}$ is a nonlinear phase-shift which
characterizes the optomechanical coupling between the resonator
and the light in the cavity. It corresponds to the normalized
phase-shift of the cavity induced by the static recoil effect of
the resonator \cite{Dorsel} and is given by:
\begin{equation}
\varphi_{\mathrm{NL}}=\frac{8\pi}{\lambda\gamma
c}\frac{P}{M\Omega_{\mathrm{m}}^2}. \label{eq:PhiNL}
 \end{equation}

At thermodynamical equilibrium, the equation of motion of the
resonator is :
\begin{equation}
x\left[\Omega\right]=\chi_{\mathrm{m}}\left[\Omega\right]\left(F_{\mathrm{T}}\left[\Omega\right]+F_{\mathrm{rad}}\left[\Omega\right]\right),
\label{eq:Mouvement}
\end{equation}
where $F_{\mathrm{T}}$ is the Langevin force responsible for the
Brownian motion \cite{Kubo}. This equation of motion can be
rewritten from (\ref{eq:Frad}) without the radiation pressure
term, but with an effective mechanical susceptibility
$\chi_{\mathrm{eff}}\left[\Omega\right]$:
\begin{equation}
{\chi_{\mathrm{eff}}}\left[\Omega\right]^{-1}={\chi_{\mathrm{m}}}\left[\Omega\right]^{-1}
+2\frac{\varphi\varphi_{\mathrm{NL}}}{\Delta}
M\Omega_{\mathrm{m}}^2.
\end{equation}
In the limit of a mechanical quality factor
$Q=\Omega_{\mathrm{m}}/\Gamma_{\mathrm{m}}\gg 1$,
$\chi_{\mathrm{eff}}\left[\Omega\right]$ still has a lorentzian
shape, but with
 effective resonance frequency and damping given by:
 \begin{eqnarray}
\Omega_{\rm
eff}&=&\Omega_{\mathrm{m}}\left(1+\,\mathrm{Re}\frac{\varphi\varphi_{\mathrm{NL}}}{\Delta}
\right)\label{eq:Omega_eff}\\
\Gamma_{\rm
eff}&=&\Gamma_{\mathrm{m}}\left(1-2Q\,\mathrm{Im}\frac{\varphi\varphi_{\mathrm{NL}}}{\Delta}
\right),\label{eq:Gamma_eff}
\end{eqnarray}
where $\Delta$ is now evaluated at the resonance frequency
$\Omega_{\mathrm{m}}$. The $Q$ factor in eq. (\ref{eq:Gamma_eff})
indicates a much stronger dependence of the damping upon radiation
pressure effects as long as the imaginary part of $1/\Delta$ stays
comparable to its real part, that is for
$\Omega_{\mathrm{m}}\simeq\Omega_{\mathrm{c}}$. The radiation
pressure effects can then increase or decrease the damping of the
resonator, depending on the sign of the detuning. Since the
Langevin force is not modified, one gets a situation somewhat
similar to the one obtained by cold damping \cite{LaserCool}: the
system still obeys the fluctuation-dissipation theorem
\cite{Kubo}, but at a different effective temperature, given for
small frequency shifts
$(\Omega_{\mathrm{eff}}-\Omega_{\mathrm{m}})$ by:
\begin{equation}
\frac{T_{\mathrm{eff}}}{T}\simeq\frac{\Gamma_{\mathrm{m}}}{\Gamma_{\mathrm{eff}}}.\label{eq:Temperature}
\end{equation}
Both radiation pressure cooling and heating can therefore be
performed, depending on the sign of the cavity detuning. A similar
result has recently been demonstrated
 with a silicon microlever, using the photothermal force rather than radiation pressure
\cite{Karrai}.

\begin{figure}
\includegraphics[width=8cm]{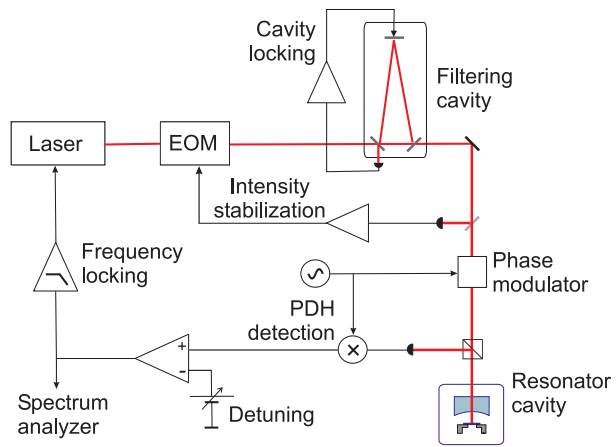}
\caption{Experimental setup used to monitor and to cool the
micro-mechanical oscillator. A Nd:YAG laser ($\lambda=1,064$ nm)
is intensity-stabilized at low frequency with an electro-optic
modulator (EOM) and spatially filtered before entering the
micro-resonator cavity. The displacement signal is extracted by
means of a Pound-Drever-Hall (PDH) phase modulation scheme using a
resonant electro-optic phase modulator. The low-frequency part of
the signal is used to lock the laser frequency around the cavity
resonance, with a preset detuning given by a DC offset. The
high-frequency part is used to monitor the resonator
displacements.} \label{fig:SetupMesure}
\end{figure}

In our experiment (Fig. 2), a silicon doubly-clamped
$(1\,\mathrm{mm}\times 1\,\mathrm{mm} \times60\,\mu\mathrm{m})$
beam with a mirror coated upon its surface is used as back mirror
of a single-ended Fabry-Perot cavity \cite{Article1}. We use one
particular  vibration mode, which has the following
characteristics: $\Omega_{\mathrm{m}}/2\pi\simeq
814\,\mathrm{kHz}$, an effective mass $M=190\,\mu\mathrm{g}$, a
spring constant $k\simeq 5\times
10^6\,\mathrm{N}.\mathrm{m}^{-1}$,
 and a mechanical quality factor $Q=10,000$ in vacuum (residual pressure
below $10^{-2}\,\mathrm{mbar}$), with the whole vacuum chamber at
room temperature $T=300\,\mathrm{K}$. Both the quality of the
low-loss dielectric coating and the low roughness
 of the resonator have allowed us to reach an optical finesse $\mathcal{F}=\pi/\gamma
=30,000$, with a cavity bandwidth
$\Omega_{\mathrm{c}}/2\pi=1.05\,\mathrm{MHz}$
($L=2.4\,\mathrm{mm}$). Such a high finesse dramatically increases
both the radiation pressure cooling effect and the sensitivity of
the phase of the reflected field to the resonator displacements.
Using a highly stabilized laser source and a Pound-Drever-Hall
(PDH) phase modulation scheme has indeed allowed us to reach a
sensitivity of $4\times10^{-19}\,\mathrm{m}/\sqrt\mathrm{Hz}$ near
the resonance frequency: the thermal peak of the resonator at room
temperature is observed with a 50 dB signal-to-noise ratio with
respect to the background thermal noise of the surrounding
vibration modes.

The calibration of our setup is performed in two steps. First, the
calibration is performed when the cavity is at resonance with a
frequency modulation of the laser \cite{EurophysLetters,Article1}.
Second, we have to take into account the lowering of the
sensitivity for a non-zero detuning, which stems from two origins:
the lower dependence of the phase response with cavity detuning
away from resonance, and the distorsion of the PDH signal for
large detunings. For that purpose, we drive the resonator into
motion (with an amplitude $\simeq10^{-13}\,\mathrm{m}$) with a
modulated electrostatic force at $814\,\mathrm{kHz}$ and monitor
the displacement signal for different detunings, and an incident
power ($50\,\mu\mathrm{W}$) low enough to insure that radiation
pressure has no effect. The observed modulation has been used to
calibrate every spectrum observed.

\begin{figure*}
\includegraphics[width=15cm] {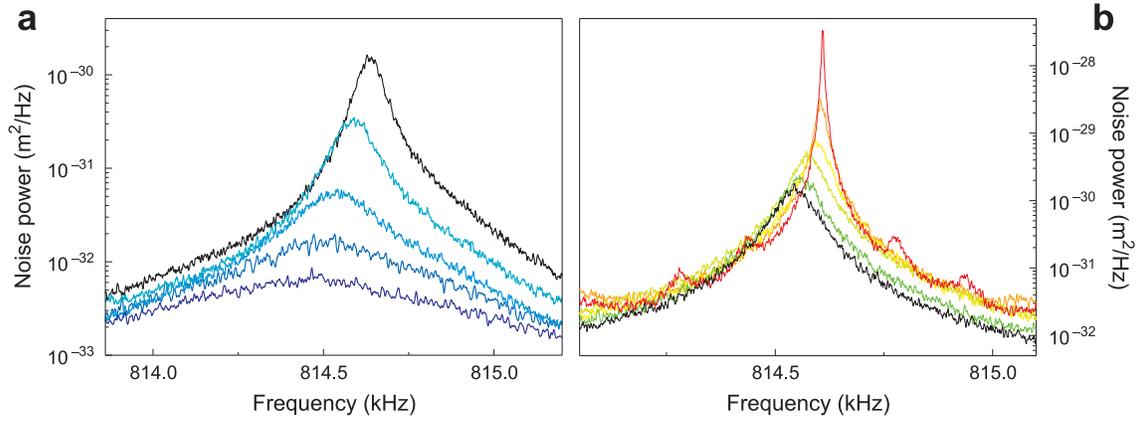}
\caption{Thermal noise spectra, normalized as micro-resonator
displacements. Black curves correspond to $\varphi=0$. {\bf a},
Curves green to blue are obtained for negative detunings
$\varphi=-0.1, \,-0.25, \,-0.4$, and $-0.6$, respectively, and for
an incident power of 5 mW. {\bf b}, Curves green to red are
obtained for positive detunings $\varphi=0.03,\,0.06,\, 0.09,\,
0.11$, and $0.13$, respectively, and for an incident power of 2.5
mW. The cooling and heating are evident through the area reduction
or increase of these spectra. Note the related drift of the
resonance frequency.} \label{fig:spectres}
\end{figure*}

Fig. 3 shows the thermal noise spectra obtained for negative
detunings and a 5 mW incident beam (a), and for positive detunings
and a 2.5 mW incident beam (b). In both cases, the black curve
corresponds to the resonant situation $\varphi=0$.  This thermal
noise spectrum (rms amplitude
$\simeq2.9\times10^{-14}\,\mathrm{m}$, driven by a
$8\times10^{-25}\,\mathrm{N}^2/\mathrm{Hz}$ Langevin force
spectral density) has been used to infer the value of the
effective mass (190 $\mu$g) used throughout the paper. This value
is in good agreement with a finite-element method (FEM)
computation, which yields a value of 130 $\mu$g \cite{Article1}.
The discrepancy is accounted for by the overall accuracy of both
the FEM computation and the fabrication of the micro-resonator, by
the imperfect overlap of the vibration mode with a non-centered
optical beam, and by the coupling of the mode with the ones of the
wafer the resonator is engraved onto.

Both  cooling and heating are evident on the noise spectra of Fig.
3. For a negative detuning, the spectra are both widened and
drastically decreased at their resonance frequencies. The decrease
of the area of the curves, which is directly related to the
effective temperature by the equipartition theorem, is a strong
indication of the temperature reduction. The situation is opposite
for positive detunings, where the spectra are strongly narrowed
and increased at resonance. The absence of any spurious noise in
the spectra is due to both the high sensitivity of our experiment
and the stable operation of the cavity. Note also in both cases
the clear frequency shift
$(\Omega_{\mathrm{eff}}-\Omega_{\mathrm{m}})$ of the resonances.

\begin{figure*}
\includegraphics[width=15cm] {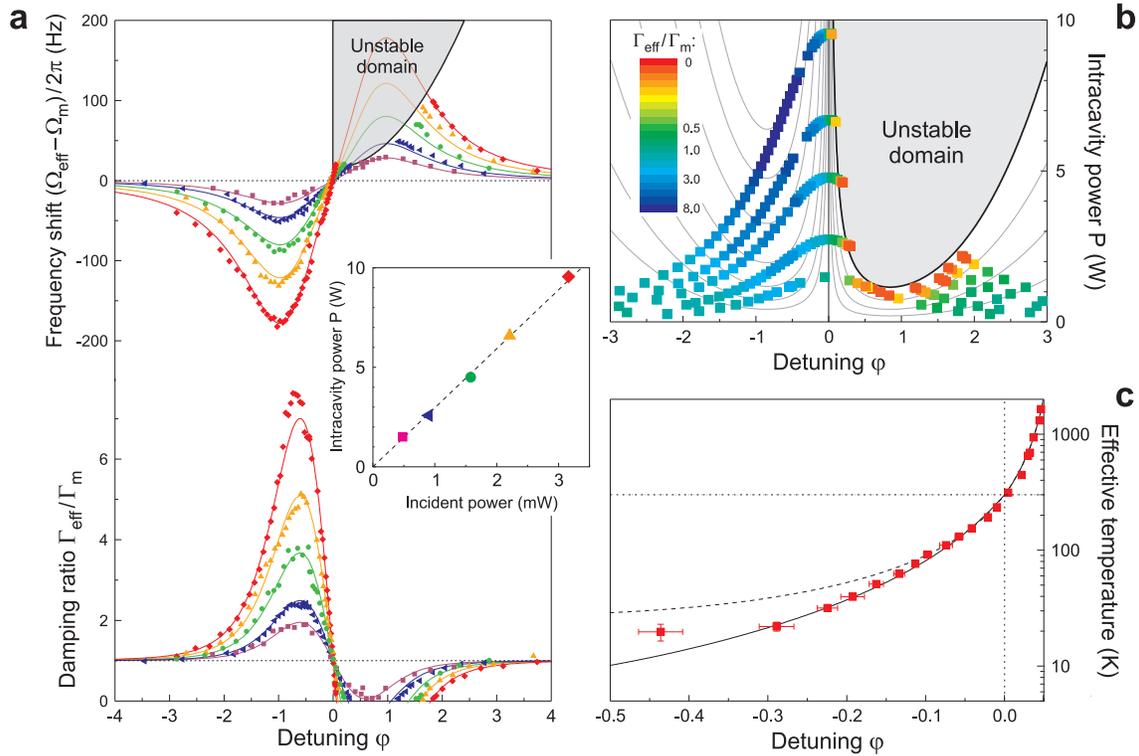}
\caption{Evolution of the cavity cooling and heating effects with
respect to the detuning $\varphi$. {\bf a}, Frequency shift
$(\Omega_{\mathrm{eff}}-\Omega_{\mathrm{m}})/2\pi$ (top) and
damping ratio $\Gamma_{\mathrm{eff}}/\Gamma_{\mathrm{m}}$
(bottom), for five values of incident power: 0.5 mW (purple), 0.9
mW (blue), 1.6 mW (green), 2.2 mW (yellow) and 3.2 mW (red).
Points are experimental results and full lines are fits obtained
from eqs. (\ref{eq:PhiNL}), (\ref{eq:Omega_eff}) and
(\ref{eq:Gamma_eff}), by adjusting the intracavity power at
resonance $P$ for each curve. Insert: evolution of the adjusted
intracavity power $P$ with the incident power. The dashed line is
the linear dependence expected  from cavity and insertion losses,
with no adjustable parameter. The shaded area in the upper curve
shows the instability zone where $\Gamma_{\mathrm{eff}}$ vanishes.
{\bf b}, Color-chart of the evolution of the damping ratio
$\Gamma_{\mathrm{eff}}/\Gamma_{\mathrm{m}}$ in the
detuning/intracavity power plane  $\{\varphi, P\}$, for the same
measurements as in {\bf a}. The value is color-coded from dark
blue (large damping, low temperature) to red (low damping, high
temperature). Gray curves are equal effective damping loci. Note
the green points (unity damping ratio) at the resonance
$\varphi=0$ and the vicinity of the red points with the
instability region (shaded area). {\bf c}, Evolution of the
effective temperature with the cavity detuning, for a 3.2 mW
incident beam. Squares: experimental points with error bars
(s.e.m.). The dotted line is the reference level at
$T=300\,\mathrm{K}$ ($\varphi=0$), the dashed line the fit with
the single oscillator model of eq. (\ref{eq:Temperature}), and the
full line the fit with a model taking into account the
out-of-resonance tails of other modes of the micro-resonator.}
\label{fig:GammaOmega}
\end{figure*}

Our experimental results can be further tested by looking at the
dependence of both the damping ratio
$\Gamma_{\mathrm{eff}}/\Gamma_{\mathrm{m}}$ and the frequency
shift $(\Omega_{\mathrm{eff}}-\Omega_{\mathrm{m}})/2\pi$ with
respect to the detuning $\varphi$. Fig. 4a presents our
experimental results, for five optical powers from
$0.5\,\mathrm{mW}$ to $3.2\,\mathrm{mW}$, along with the
theoretical fits deduced fom eqs. (\ref{eq:Omega_eff}) and
(\ref{eq:Gamma_eff}). Both are in very good agreement. The insert
shows the intracavity power at resonance $P$, single free
parameter derived from the preceeding fits, as a function of the
incident power $P_{\mathrm{in}}$. It exhibits the expected linear
dependence, with a slope $P/P_{\mathrm{in}}=2,970$ in agreement at
the 1\% level with the one deduced from cavity and insertion
losses. This clearly shows that the observed cooling effect is
solely due to radiation pressure. Indeed, with the estimated
mirror absorption ($\simeq$ 3 ppm), thermoelastic effects are
expected to be $10^6$ times lower. Bolometric effects
\cite{Karrai} are difficult to estimate, but both the low
absorption and the large thickness of the resonator still are in
favor of a negligible effect. We have also performed the same
experiment for other modes of the resonator, with the same
excellent agreement with theory. In particular, a mode with a
resonance frequency $\Omega_{\mathrm{m}}=2\pi\times
2,824\,\mathrm{kHz}$ larger than the cavity bandwidth
$\Omega_{\mathrm{c}}$ exhibits a frequency shift
$(\Omega_{\mathrm{eff}}-\Omega_{\mathrm{m}})$ with an opposite
sign, as expected from eqs. (\ref{eq:Delta}) and
(\ref{eq:Omega_eff}) for not too large detunings $\varphi$.

Figure 4b presents a color-chart of the dependence of the
effective damping $\Gamma_{\mathrm{eff}}$ in the
detuning/intracavity power plane. Experimental series of points
were taken for fixed stabilized incident powers, following the
Airy curves in the plane. The color code ranges from dark blue
(for large $\Gamma_{\mathrm{eff}}$ and low
$\mathrm{T}_{\mathrm{eff}}$) to red (for low
$\Gamma_{\mathrm{eff}}$ and large $\mathrm{T}_{\mathrm{eff}}$).
The green points (corresponding to $\Gamma_{\mathrm{eff}}\simeq
\Gamma_{\mathrm{m}}$) are located at the resonance, at large
detunings (where the slope of the Airy peak vanishes) and at low
intracavity power (where radiation pressure effects are weaker).
The highest cooling (dark blue) is obtained near a negative
detuning $\varphi\simeq -0.6$ and for a high intracavity power
$P$, whereas the largest heating effect (red) is obtained for a
positive detuning. Gray curves correspond to equal effective
dampings $\Gamma_{\mathrm{eff}}$ and the black one
 to $\Gamma_{\mathrm{eff}}=0$. Above this line, the
resonator becomes unstable and starts to oscillate at its
effective resonance frequency, as has already been demonstrated
with a micro-toroidal resonator \cite{Vahala}. We have observed a
similar effect, as shown on Figs. 4a and 4b, where the instability
region is displayed with no adjustable parameter. Also note that
the secondary peaks on the red spectrum of Fig. 3b (more than 50
dB below the resonance spectrum level) are due to the lower
stability of the cavity operation close to the instability region.
This is to our knowledge the first experimental demonstration of
such a radiation-pressure instability in an open optical cavity.

Fig. 4c displays the variations of the effective temperature with
the detuning $\varphi$. For a fixed input power of
$3.2\,\mathrm{mW}$, $T_\mathrm{eff}$ ranges from $20\,\mathrm{K}$
at an optimum detuning $\varphi\simeq-0.45$, to
$2\,000\,\mathrm{K}$ for a positive detuning, very close to the
instability region. The dependence with $\varphi$ is well
accounted for by our single oscillator theoretical model (eq.
\ref{eq:Temperature}) for positive and negative but not too large
detunings (dashed line). For larger negative detunings, the noise
spectrum is so low that the thermal noise background due to other
vibration modes of the resonator is no longer negligible: a slight
modification of our model taking this background into account well
depicts the characteristics of our experimental results (full
line). Due to the low absorption of the micro-resonator, the
residual discrepancy at large detuning (left point in Fig. 4c)
cannot be related to a residual heating process. Furthermore,
effective temperatures as low as 10 K have been obtained with a 12
mW incident beam.

Our experimental setup is the first high-sensitivity optical
displacement sensor to show a direct effect of radiation pressure,
an ubiquitous -though extremely weak- effect that every optical
experiment will eventually be sensitive to. Further enhancement
and low temperature operation of such a system may lead to the
experimental demonstration of the quantum ground state of a
mechanical resonator
\cite{Roukes,Cleland,Schwab,Karrai,Article1,SpieAustin}:
radiation-pressure cooling is a fundamental process which
eventually involves nothing but a mechanical oscillator in
interaction with one cavity mode and its quantum fluctuations,
yielding a reliable estimation of the cooling limit. Furthermore,
our knowledge of the other modes' dynamical behavior
\cite{Article1} appears as a major issue on the road to the
quantum ground state as their out-of-resonance tails will no
longer be negligible for extreme cooling ratios. Other possible
novel effects in quantum optics include the experimental
demonstration of the Standard Quantum Limit
\cite{Caves81,Jaekel90}, radiation-pressure induced squeezing of
light \cite{Fabre94}, Quantum Non Demolition measurements
\cite{Heidmann97} or non-classical states of the resonator motion
\cite{Bose97,EPLEntanglement}.

\bigskip
 Acknowledgements are due to L. Rousseau for the
fabrication of the micro-resonator and to J.-M. Mackowski and his
group for the optical coating of the resonator. This work was
partially funded by EGO (collaboration convention EGO-DIR-150/2003
for a study of quantum noises in gravitational wave
interferometers). Laboratoire Kastler Brossel is "Unit\'{e} mixte
de recherche du Centre National de la Recherche Scientifique, de
l'Ecole Normale Sup\'{e}rieure et de l'Universit\'{e} Pierre et
Marie Curie".


\end{document}